\documentclass[aps, prb, twocolumn]{revtex4-2}

\usepackage{amsmath,amsfonts,amssymb}
\usepackage{graphicx,color}
\usepackage{hyperref}
\usepackage{epstopdf}
\usepackage{float}
\usepackage{multirow}
\newcommand{\ket}[1]{\left|#1\right\rangle}
\newcommand{\bra}[1]{\left\langle#1\right|}

\newcommand{\icol}[1]{\left(\begin{smallmatrix}#1\end{smallmatrix}\right)} 

\begin{document}

\title{Deterministic Grover search with a restricted oracle}

\author{Tanay Roy$^{1}$}
\email{Corresponding author: roytanay@uchicago.edu}
\author{Liang Jiang$^{2}$, David I. Schuster$^{1}$}

\affiliation{$^{1}$Department of Physics and James Franck Institute, University of Chicago, Chicago, Illinois 60637, USA}
\affiliation{$^{2}$Pritzker School of Molecular Engineering, University of Chicago, Chicago, Illinois 60637, USA}

\date{\today}

\begin{abstract}
Grover's quantum search algorithm provides a quadratic quantum advantage over classical algorithms across a broad class of unstructured search problems. The original protocol is probabilistic, returning the desired result with significant probability on each query, but in general, requiring several iterations of the algorithm. We present a modified version to return the correct result with certainty without having user control over the quantum search oracle. Our deterministic, two-parameter ``D2p'' protocol utilizes generalized phase rotations replacing the phase inversions after a standard oracle query. The D2p protocol achieves a 100\% success rate in no more than one additional iteration compared to the optimal number of steps as in the original Grover's search enabling the same quadratic speed up. We also provide a visualization using the Bloch sphere for enhanced geometric intuition.

\end{abstract}

\maketitle

\section{Introduction}

An efficient algorithm to search a large unstructured search space has a wide range of applications. While the time complexity of a classical search algorithm scales linearly with the size of the space, Grover's quantum search algorithm~\cite{Grover_algo} provides a quadratic speedup. Though the quantum advantage is not exponential as in some quantum algorithms it can be applied very broadly, to any problem whose result can be verified efficiently.  The search is performed by an ``oracle'', a quantum algorithm that the user may not have access to.  The oracle could be a quantum random access memory~\cite{Qram2008, Qram_architecture2008} to access an unstructured classical or quantum database. Alternatively, the oracle could be a quantum version of a one-way function such as a hash, symmetric key encryption, or number theory conjecture, etc.  Given a space with $N$ unsorted inputs and quantum oracle that identifies $M$ marked states, Grover's search algorithm is guaranteed to produce better than 50\% success probability with $\mathcal{O}(\sqrt{N/M}) $ oracle queries, whereas a classical algorithm needs on average $N/(2M)$ interrogations.

Grover's algorithm is composed of two steps --- the oracle query, which flips the phase of the marked states, and the application of the diffusion operator (also known as the reflection or inversion operator) that amplifies the amplitude of the marked states. In the original protocol~\cite{Grover_algo}, both steps use a phase-flip operator that restricts the evolution of the initial superposition state in a way that the success is probabilistic. A family of non-deterministic Grover-type searching algorithms has also been found to provide similar quadratic speedup~\cite{GroverFamily2000}. One can achieve the target state with certainty by controlling the phases of the phase-flip and diffusion operators when the ratio $\lambda={M/N}$ is known~\cite{Long2001certain_Grover, Toyama2013certain_Grover}. Other works aim to improve the success rate for an unknown $\lambda$ (with a modest guess of the lower bound) by performing multi-phase matching~\cite{toyama2008multiphase, Chuang2014fixed_point, Li2007phase_matching}. However, these protocols require one to control the phase of the oracle, which might not always be plausible as the user may not have knowledge of or access to the oracle. Practically, the search oracle should be treated as a fixed unitary determined by some physical process with no user-tunable parameters.

In this work, we present an algorithm to find a marked state \textit{deterministically} with the constraint that the user does not have control over the oracle phase. We show that only two phase parameters for the consecutive diffusion operators are sufficient to find a target state with certainty by making $k_{\rm opt}=\left \lceil \frac{\pi}{4\sin^{-1}\sqrt{\lambda}} -\frac{1}{2}\right \rceil $ oracle queries for a given $\lambda$.

\begin{figure*}[t]
	\centering\includegraphics[width=\textwidth]{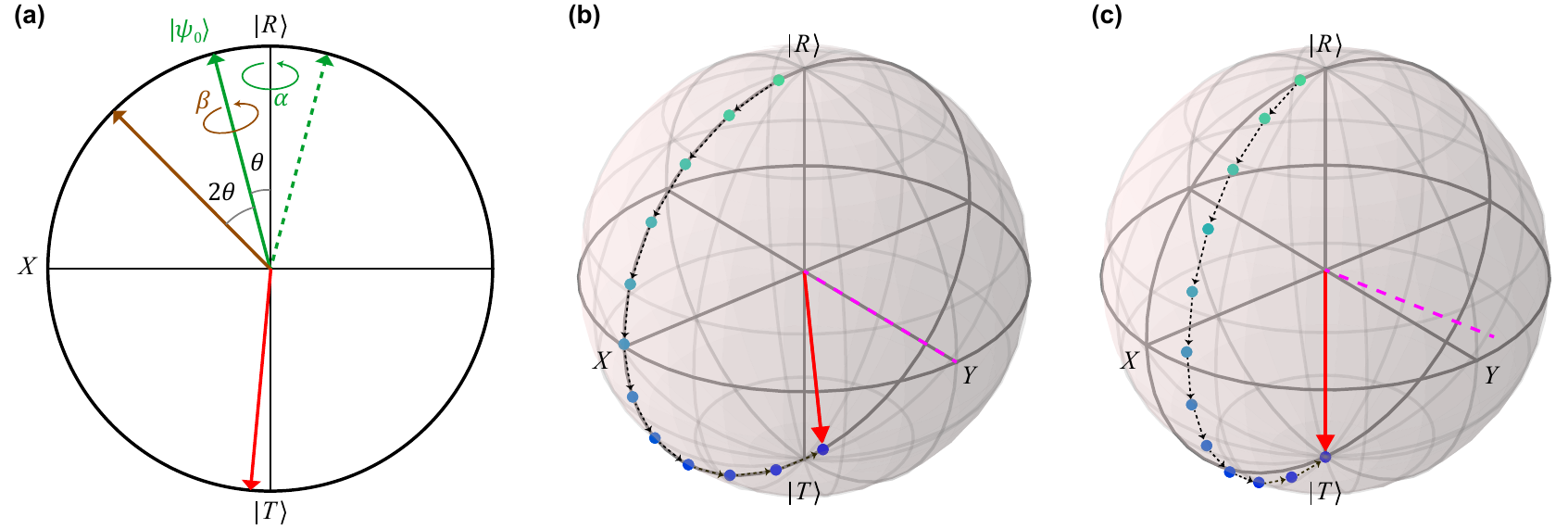}
	\caption{Trajectory of the state vector on a Bloch sphere spanned by the equal-superposition of marked states $\ket{T}$ (south pole) and unmarked states $\ket{R}$ (north pole). (a) The initial equal superposition state $\ket{\psi_0}$ (solid green arrow) makes a polar angle $\theta$ determined by the ratio of marked to unmarked state counts $\lambda$. The plane containing the vectors $\ket{\psi_0}$ and $\ket{R}$ is assumed to be the $ZX$ plane of the Bloch sphere. The generalized oracle operator $S_0(\alpha)$ (see Eq.~\eqref{eq:oracle}) rotates the state vector about $z$-axis by an angle $\alpha$. Similarly, the generalized reflection operator $S_r(\beta)$ (see Eq.~\eqref{eq:reflect}) performs a rotation of the state vector about the direction of $\ket{\psi_0}$ by an angle $\beta$. The oracle in the original Grover's algorithm with $\alpha=\pm\pi$ flips the phase of $\ket{T}$ resulting in the dashed green vector and then the reflection operator (with $\beta=\pm \pi$) inverts the state with respect to $\ket{\psi_0}$ leading to the brown vector when starting from $\ket{\psi_0}$. As a result, a single Grover iterate effectively rotates the state vector by $2\theta$ about the $y$-axis in each iteration. (b) Consequently, the trajectory of the state vector is always confined in the $ZX$ plane, perpendicular to the $y$-axis (dashed magenta line) in the original Grover's algorithm and the final state (red arrow) doesn't always end up along the south pole after an integer number of steps. (c) The protocol in Ref.~\onlinecite{Long2001certain_Grover} achieves zero theoretical failure rate by rotating the state vector in a plane such that it always lands along the south pole. The corresponding axis of rotation (dashed magenta line) is carefully chosen and doesn't lie along the $y$-axis in general. In this case, the user is assumed to have control over both the oracle and the reflection operator steps so that one can set $\alpha=\beta=\theta_0$ (see Eq.~\eqref{eq:theta0}).}
	\label{fig:traj}
\end{figure*}

\section{Overview of Grover's algorithm}
The original Grover's algorithm constitutes of successive application of the oracle and diffusion operators on the initial equal-superposition state (containing mostly unmarked states) that is transformed into a  superposition of mostly marked states. Assuming $N=2^n$ as the number of total states, one can prepare an equal-superposition state $\ket{\psi_0}$ by applying a Walsh-Hadamard transformation individually to all the qubits initiated to $\ket{0}$. Instead of using the full $2^n$-dimensional Hilbert space, it is more convenient to map the system to a two-dimensional sub-space spanned by the orthogonal vectors $\ket{T}$ and $\ket{R}$, where $\ket{T} (\ket{R})$ represents the equal-superposition of all marked (unmarked) states  $\ket{t_j} (\ket{r_j})$,
\begin{subequations}
	\begin{align}
	\ket{T} &= \dfrac{1}{\sqrt{M}} \sum_{j=1}^{M}\ket{t_j}, \\
	\ket{R} &= \dfrac{1}{\sqrt{N-M}} \sum_{j=1}^{N-M}\ket{r_j}.
	\end{align}
\end{subequations}
Then the initial state can be expressed in the new basis $\ket{R}=\icol{1\\0}$ and $\ket{T}=\icol{0\\1}$ as
\begin{eqnarray}
\label{eq:psi_0}
	\ket{\psi_0}=\sqrt{1-\lambda} \ket{R} + \sqrt{\lambda} \ket{T} = 
	\begin{pmatrix}
		\sqrt{1-\lambda} \\
		\sqrt{\lambda}
	\end{pmatrix}.
\end{eqnarray}
The evolution of $\ket{\psi_0}$ can be visualized as a moving unit vector on the Bloch sphere spanned by $\ket{R}$ (north pole) and $\ket{T}$ (south pole) as shown in Fig.~\ref{fig:traj}(a). The initial state (solid green arrow) lies in the $ZX$ plane making an angle $\theta=2 \sin^{-1}\sqrt{\lambda}$ with the $z$-axis. The oracle is a unitary operator
\begin{equation}
S_o(\alpha) = I - (1-e^{i\alpha})\ket{T}\bra{T} = 
\begin{pmatrix}
1 &0  \\
0 &e^{i \alpha}
\end{pmatrix}
\label{eq:oracle}
\end{equation}
representing a generalized controlled-phase gate --- rotation of the vector about $z$-axis by an angle $\alpha$ with $I$ being the identity operator. The special case of $S_o(\pi)$ on $\ket{\psi_0}$ can be simply thought as a reflection with respect to the vertical axis.

\begin{figure*}[t]
	\centering\includegraphics[width=\textwidth]{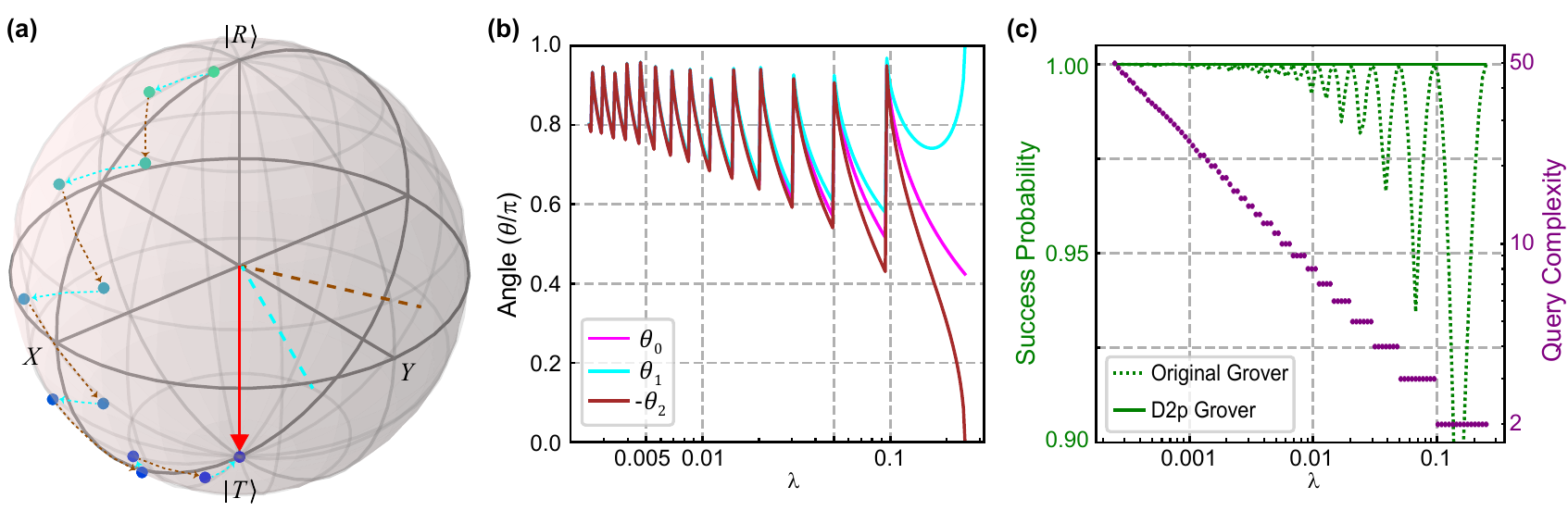}
	\caption{The D2p Grover's search protocol. (a) The state vector is rotated about two axes (cyan and brown dashed lines) alternatively by amounts $\theta_1$ and $\theta_2$. One can always find parameters $\{\theta_1,\theta_2\}$ such that the final state (red arrow) ends up along the south pole providing zero failure rate for $\lambda \le 1/4$ when the number of iterations $k\ge k_{\rm opt}$. (b) The values of $\theta_1$ (cyan curve) and $\theta_2$ (brown curve) are plotted as a function of $\lambda$ obtained by numerically solving Eq.~\eqref{eq:thetas} when $k_{\rm opt}$ is even and a similar equation when $k_{\rm opt}$ is odd (see Appendix~\ref{app:odd_eq}). A plot of $\theta_0$ is also shown for reference. (c) Comparison of success rate between the original Grover's search (dashed green curve) and our D2p protocol (solid green line). The corresponding query complexity $k_{\rm opt}$ of the D2p protocol is shown (purple curve) which is at maximum one extra step compared to the optimal number of steps $k'_{\rm opt}$ for the original Grover's search.}
	\label{fig:dg}
\end{figure*}

Next, we define the generalized Grover's reflection following the convention used in Ref.~\onlinecite{Chuang2014fixed_point} (up to a global phase)
\begin{equation}
\begin{split}
S_r(\beta) &= e^{i\beta}\left( I - (1 - e^{-i\beta})\ket{\psi_0}\bra{\psi_0}\right) \\
&=
\begin{pmatrix}
1-(1-e^{i\beta})\lambda & (1 - e^{i\beta})\sqrt{\lambda (1-\lambda)} \\
(1 - e^{i\beta})\sqrt{\lambda (1-\lambda)} & 1-(1-e^{i\beta})(1-\lambda)
\end{pmatrix},
\end{split}
\label{eq:reflect}
\end{equation}
which represents a rotation of the state about $\ket{\psi_0}$ by an angle $\beta$ (see Fig.~\ref{fig:traj}(a)). The product of the oracle and the reflection operator is often called the Grover's iterate $G(\alpha,\beta)=-S_r(\beta)S_o(\alpha)$. The original Grover's iterate with $\alpha=\pm \pi$ and $\beta=\pm \pi$  rotates the state vector by $2\theta$ about the $y$-axis (dashed pink line in Fig.~\ref{fig:traj}(b)), restricting the trajectory in the $ZX$ plane. Since the angular distance of $\ket{\psi_0}$ from the south pole is $\pi-\theta$, the number of steps needed to reach $\ket{T}$ becomes $(\pi-\theta)/(2\theta)$, which is, in general, a fractional number. Therefore, the optimal number of steps for the original Grover's search becomes the nearest integer~\cite{Long2001certain_Grover}
\begin{equation}
k'_{\rm opt} = \left \lfloor \dfrac{\pi}{2\theta} - \dfrac{1}{2} \right \rceil = \left \lfloor \dfrac{\pi}{4\sin^{-1}\sqrt{\lambda}} - \dfrac{1}{2} \right \rceil.
\end{equation}
In general the final state vector (red arrow in Fig.~\ref{fig:traj} (b)) will not always align with south pole providing a maximum success probability of $\sin[(k'_{\rm opt}+1/2)\theta]^2$. This situation of undershooting (overshooting) the target state is often called ``undercooking (overcooking)". One can, however, cleverly choose a different plane of rotation so that the final state always lands on the south pole in $k \ge k_{\rm opt}$ steps as shown in Fig.~\ref{fig:traj} (c). The corresponding Grover's iterate needs $\alpha=\beta=\theta_0$ with~\cite{Long2001certain_Grover}
\begin{equation}
	\theta_0 = 2 \sin^{-1}\left( \dfrac{1}{\sqrt{\lambda}} \sin\left( \dfrac{\pi}{4k+2} \right) \right),
	\label{eq:theta0}
\end{equation}
which corresponds to an axis of rotation (dashed pink line in Fig.~\ref{fig:traj}(c)) not parallel to the $y$-axis in general.

\section{The D2p protocol}
The success of the protocol in Ref.~\onlinecite{Long2001certain_Grover} relies on the fact that the oracle is user-controllable which might not be always feasible. In this paper, we explore the possibility of deterministic outcome using a fixed oracle. In particular, we consider the standard phase flip operator with $\alpha=\pi$. We show that, interestingly enough, only two phase parameters are needed to obtain zero failure rate, i.e., we apply Grover iterates $G(\pi,\theta_1)= G_d(\theta_1)$ and $G(\pi,\theta_2)= G_d(\theta_2)$ alternatively to the initial state $\ket{\psi_0}$ until $k_{\rm opt}$ oracle queries are made. We call it deterministic 2-parameter (D2p) Grover's search algorithm. The requirement of only two phase parameters can be very intuitively understood from the fact that one needs only two non-colinear axes of rotation to span the full $SU(2)$ space. An example of the resulting trajectory for $\lambda=0.005$ is illustrated in Fig.~\ref{fig:dg} (a).

\begin{figure*}[t]
	\centering\includegraphics[width=\textwidth]{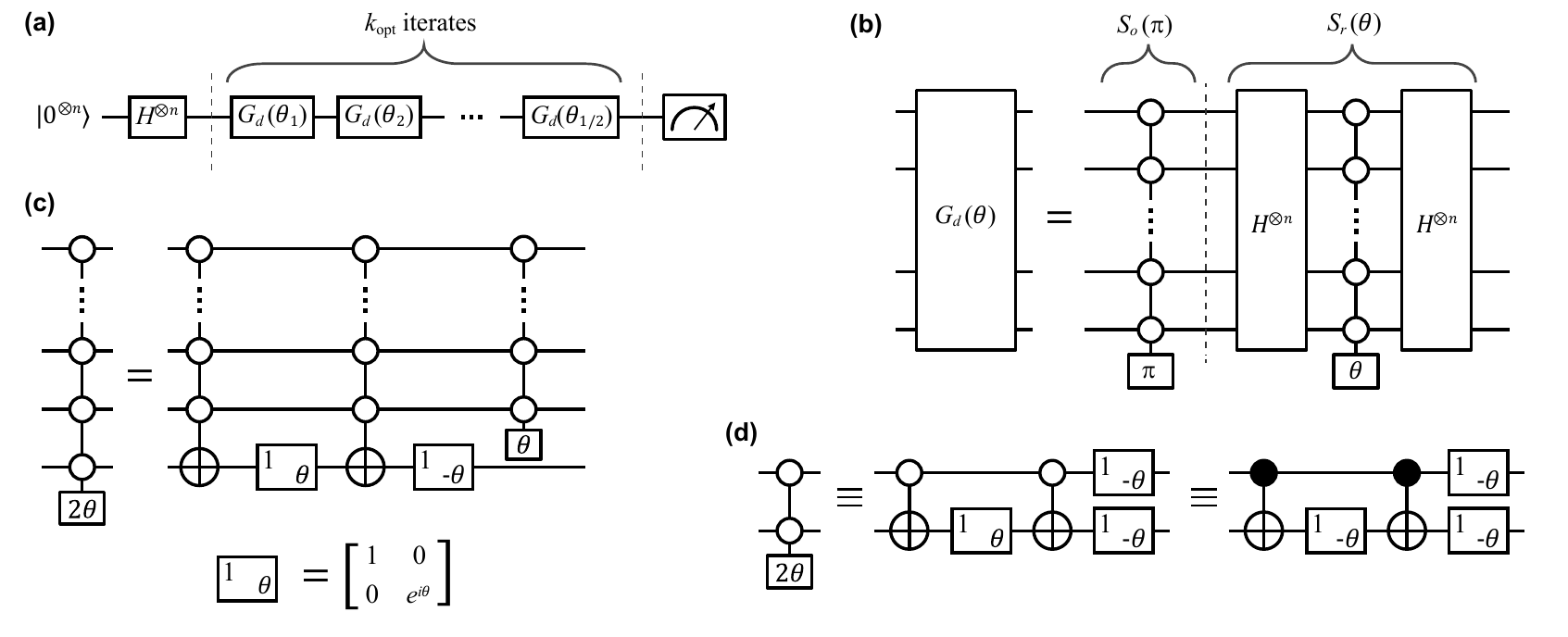}
	\caption{Circuit implementation of the D2p Grover's search protocol without any ancillary qubit. (a) A schematic of the quantum circuit showing alternate application of the Grover iterate $G_d(\theta)$ on the equal-superposition state $\ket{\psi_0}$ obtained by applying Walsh-Hadamard transformation on individual qubits initialized to $\ket{0}$. After $k_{\rm opt}$ iterations a final projective measurement is performed to retrieve one of the marked states with certainty. (b) Decomposition of one Grover iterate using generalized multiply-controlled phase gate and Walsh-Hadamard gates. It consists of the standard oracle operator $S_o(\pi)$ and the generalized reflection operator $S_r(\theta)$. (c) A method to construct the generalized multiply-controlled phase gate using multiply-controlled NOT gates and single-qubit $Z$-rotations. The matrix at the bottom panel represents shifting phase of the $\ket{1}$ component by $\theta$. (d) Break down of the generalized two-qubit phase gate using CNOT gates single-qubit rotations up to a global phase.}
	\label{fig:circuit}
\end{figure*}

The goal of this protocol then reduces to determining the phases that will ensure landing of the final state along the south pole. We first consider the case when the number of queries $k$ is an even number, so that the final state is
\begin{equation}
	\ket{\psi_f} = \Big(G_d(\theta_2)\cdot G_d(\theta_1)\Big)^{k/2}\ket{\psi_0}.
\end{equation}
Imposing the condition $\langle R\ket{\psi_f}=0$ leads to the following two equations
\begin{subequations}
	\begin{align}
	1 + 4\lambda (1-2\lambda) \sin\left( \dfrac{\theta_1}{2} \right) \sin\left( \dfrac{\theta_2}{2} \right) \dfrac{\tan(\frac{k}{2} \phi)}{\sin(\phi)} &= 0, \\
	(1-4\lambda) \tan\left( \dfrac{\theta_1}{2} \right) + \tan\left( \dfrac{\theta_2}{2} \right) &= 0,
	\end{align}
\label{eq:thetas}
\end{subequations}
where
\begin{equation}
	\cos(\phi) = \cos\left( \dfrac{\theta_1+\theta_2}{2} \right) + 8\lambda(1-\lambda) \sin\left( \dfrac{\theta_1}{2} \right) \sin\left( \dfrac{\theta_2}{2} \right).
\end{equation}
These equations can always be solved for $\{\theta_1,\theta_2\}$ when $k\ge k_{\rm opt}$ and $\lambda\le 1/4$.

When $k$ is odd, the final state is 
\begin{equation}
\ket{\psi_f} = G_d(\theta_1) \cdot \Big(G_d(\theta_2)\cdot G_d(\theta_1)\Big)^{\lfloor k/2 \rfloor}\ket{\psi_0},
\end{equation}
and one can find two equations similar to Eq.~\eqref{eq:thetas} that can be solved to obtain the optimal phase parameters (see Appendix~\ref{app:odd_eq} for explicit equations). Figure \ref{fig:dg} (b) plots $\theta_0, \theta_1$ and $\theta_2$ as a function of $\lambda$ with $k=k_{\rm opt}$. The sharp jumps occur when the query complexity $k_{\rm opt}$ changes by one as depicted in Fig.~\ref{fig:dg}(c). Note that for sufficiently small $\lambda$, $\theta_1\approxeq-\theta_2 \approxeq \theta_0$. A comparison of the success probabilities between the standard Grover's search (green dashed line) and the D2p protocol (solid green line) is also shown in Fig.~\ref{fig:dg}(c). One can always choose more than two phase parameters to obtain determinism, but any additional phase doesn't provide extra degree of freedom (as the reduced Bloch sphere has only two dimensions) and thus reaching the target state faster than $k_{\rm opt}$ steps is not possible.

The D2p protocol can be generalized to quantum amplitude amplification~\cite{Hoyer1997amp_amplification, Grover1998amp_amplification, brassard2002amp_amplification, ambainis2004search_algo} where we prepare a random initial state $\ket{\psi'_0} = \mathcal{A}\ket{0}$ instead of the equal-superposition state $\ket{\psi_0}$. Here, $\mathcal{A}$ could be any unitary as long as $\ket{\psi'_0}$ has a finite overlap with marked state $\ket{T}$. Similar to Eq.~\eqref{eq:psi_0}, one can use the basis $\ket{T}$ and $\ket{R'}$ to express
\begin{eqnarray}
	\ket{\psi'_0}=\sqrt{1-\lambda'} \ket{R'} + \sqrt{\lambda'} \ket{T} = 
	\begin{pmatrix}
		\sqrt{1-\lambda'} \\
		\sqrt{\lambda'}
	\end{pmatrix},
\end{eqnarray}
where $\ket{R'}$ is now consists of a generic (normalized) superposition of unmarked states. The reflection unitary in Eq.~\eqref{eq:reflect} will be similarly modified to 
\begin{equation}
\begin{split}
S'_r(\beta) &= e^{i\beta}\left( I - (1 - e^{-i\beta})\ket{\psi'_0}\bra{\psi'_0}\right) .
\end{split}
\label{eq:reflect_2}
\end{equation}
An identical analysis can be performed by replacing $\lambda$ and $\ket{\psi_0}$ with $\lambda'$ and $\ket{\psi'_0}$ to arrive at Eqs.~\eqref{eq:thetas}. However, note that the knowledge of overlap $\lambda'=|\bra{\psi'_0}T\rangle|^2$ is still required to achieve determinism.

Next, we turn to the circuit implementation of the D2p protocol as depicted in Fig.~\ref{fig:circuit}(a). Hadamard gates are applied to individual qubits (initialized to $\ket{0}$) to prepare the equal-superposition state $\ket{\psi_0}$. The modified Grover's iterates $G_d(\theta)$ are applied alternatively with $\theta=\theta_1$ and $\theta=\theta_2$ for $k_{\rm opt}$ times. The last iterate is $G_d(\theta_{1(2)})$ if $k_{\rm opt}$ is an odd (even) number and the final state becomes an equal superposition of the marked states guaranteeing a success when a projective measurement is performed. Each Grover iterate $G_d(\theta)$ is composed of two generalized multiply-controlled phase gates and Hadamard gates as shown in Fig.~\ref{fig:circuit}(b). A generalized multiply-controlled phase gate involving $n$ qubits can be deconstructed using two generalized multiply-controlled NOT gates involving $n$ qubits and one generalized multiply-controlled phase gate involving $(n-1)$ qubits along with two single-qubit phase gates as displayed in Fig.~\ref{fig:circuit}(c). This decomposition can be inductively applied to construct the target gate using $\mathcal{O}(n^2)$ controlled NOT gates and single qubit rotations~\cite{saeedi2013n_toffoli}. The final two-qubit generalized controlled-phase gate in this decomposition method can be constructed using two two-qubit gates and three single-qubit $Z$ rotations as shown in Fig.~\ref{fig:circuit}(d).

\section{Conclusion}
We have presented a modified version of Grover's search algorithm to find the correct answer with zero failure rate without having user control over the oracle implementation. The main advantage of our D2p protocol is that it requires only two phase parameters to be used in the generalized multiply-controlled phase gates while providing quadratic speedup. The phases can be numerically determined for any marked-to-total number of states ratio $\lambda \le 1/4$. For $1/4<\lambda<1/2$, one can use a single query of standard Grover's search~\cite{Grover_algo} (non-deterministic) or any classical algorithm as there is no significant quantum advantage. The visual representation of this protocol using the Bloch-sphere picture makes it very intuitive and can be adapted to other phase-matching protocols~\cite{toyama2008multiphase, Chuang2014fixed_point, Li2007phase_matching}. 

The D2p protocol can be readily applied to any framework where the quantum amplitude amplification~\cite{Grover1998amp_amplification, brassard2002amp_amplification, ambainis2004search_algo, Bae2021QAAO}, a generalization of Grover's search algorithm, is used including search using qudits. A few examples include element distinctness problem~\cite{buhrman2001distinct, ambainis2007distinct}, minima finding~\cite{durr1996global_minima, aaronson2006local_minima}, and collision problems~\cite{brassard1997collision}. Other interesting directions worth pursuing would be to investigate the range of (fixed) phases in the oracle compatible with the D2p protocol (see Appendix~\ref{app:alpha}), and explore deterministic variants of the generalized Grover-type searching algorithms~\cite{GroverFamily2000}. One drawback is, however, the requirement of accurate knowledge of $\lambda$, which is true for other deterministic search algorithms as well~\cite{Long2001certain_Grover}. There are attempts to bound the failure rate when $\lambda$ is unknown by using multiple phase-matching~\cite{Chuang2014fixed_point} albeit at the expense of using more oracle queries than the standard optimal number $k_{\rm opt}$ and having control over the oracle operator. Another extension of our protocol would be to address the possibility of achieving similar fixed-point behavior without user-controlled oracles.

\section{Acknowledgements}
This work was supported by the Army Research Office under Grant No. W911NF-18-1-0125, National Science Foundation Grant No. PHY-1653820, Air Force Office of Scientific Research under Grant No. FA9550-21-1-0209, Depart of Energy Q-NEXT Center, NTT Research, and the Packard Foundation. This work was also funded in part by EPiQC, an NSF Expedition in Computing, under grant CCF1730449. We thank Dongjin Lee, Hyeokjea Kwon, Joonwoo Bae, and Saptarshi Roy Chowdhury for bringing transcription related errors to our attention.
\\ 
\noindent

\appendix

\section{Exponentiation of the Grover iterates}
If a matrix $M$ can be expressed as
\begin{equation}
    M = \cos{(\phi)} \ I + i \sin{(\phi)} \sum_{j=x,y,z} n_j \sigma_j, 
\end{equation}
where $\sigma_{x,y,z}$ are the standard Pauli matrices and $\sum_{j} |n_j|^2=1$, then $k$-th power of $M$ becomes
\begin{equation}
    M^k = \cos{(k\phi)} \ I + i \sin{(k\phi)} \sum_{j=x,y,z} n_j \sigma_j.
\end{equation}
In order to remove a global phase (inconsequential) we choose
\begin{equation}
    M=e^{-i(\theta_1 + \theta_2)/2} \ G_d(\theta_2) \cdot G_d(\theta_1),
\end{equation}
leading to
\begin{subequations}
\begin{align}
    \cos(\phi) &= \cos\left( \dfrac{\theta_1+\theta_2}{2} \right) + 8\lambda(1-\lambda) \sin\left( \dfrac{\theta_1}{2} \right) \sin\left( \dfrac{\theta_2}{2} \right), \\
    n_x &= \dfrac{2\sqrt{\lambda(1-\lambda)}}{\sin(\phi)}  \sin{\left( \dfrac{\theta_1-\theta_2}{2} \right)}, \\
    n_y &= \dfrac{4 (1-2\lambda)\sqrt{\lambda(1-\lambda)}}{\sin(\phi)}  \sin{\left( \dfrac{\theta_1}{2} \right)} \sin{\left( \dfrac{\theta_2}{2} \right)}, \\
    n_z &= -\dfrac{(1-2\lambda)}{\sin(\phi)}  \sin{\left( \dfrac{\theta_1+\theta_2}{2} \right)}.
    \end{align}
\end{subequations}
When $k_{\rm opt}$ is even, we compute the final state as
\begin{equation}
    \ket{\psi_f}=M^{k_{\rm opt}/2} \ket{\psi_0},
\end{equation}
otherwise,
\begin{equation}
    \ket{\psi_f}=G_d(\theta_1) \cdot M^{(k_{\rm opt}-1)/2} \ket{\psi_0}.
\end{equation}
Setting the real and imaginary components of $\langle R\ket{\psi_f}$ separately to zero we obtain the relevant equations for the phase parameters.

\section{Equations when $k_{\rm opt}$ is odd}
\label{app:odd_eq}
\begin{widetext}
\begin{subequations}
	\begin{align}
	\begin{split}
		2\lambda + (1-2\lambda)\cos(\theta_1) - (1-2\lambda)\sin\left( \dfrac{\theta_1}{2} \right)
		\left[ \sin(\theta_1) \cos\left( \dfrac{\theta_2}{2} \right) + \Big(1+4\lambda-8\lambda^2 + (1-8\lambda+8\lambda^2)\cos(\theta_1)\Big) \sin\left( \dfrac{\theta_2}{2} \right) \right]
		\dfrac{\tan(\frac{k-1}{2} \phi)}{\sin(\phi)} \\ = 0, 
		\end{split}
		\\
		\begin{split}
		(1-2\lambda)\sin(\theta_1) +\left[ (1-2\lambda)\left( 
		8\lambda(1-\lambda)\sin\left( \dfrac{\theta_1}{2} \right) \sin(\theta_1)\sin\left( \dfrac{\theta_2}{2} \right) +
		\cos(\theta_1)\sin\left( \dfrac{\theta_1+\theta_2}{2} \right) \right) -
		2\lambda\sin\left( \dfrac{\theta_1-\theta_2}{2} \right)\right]
		\dfrac{\tan(\frac{k-1}{2} \phi)}{\sin(\phi)} \\ = 0.
		\end{split}
	\end{align}
\end{subequations}
\end{widetext}

\section{Other oracle phases}
\label{app:alpha}
\begin{figure}[t]
	\centering\includegraphics[width=\columnwidth]{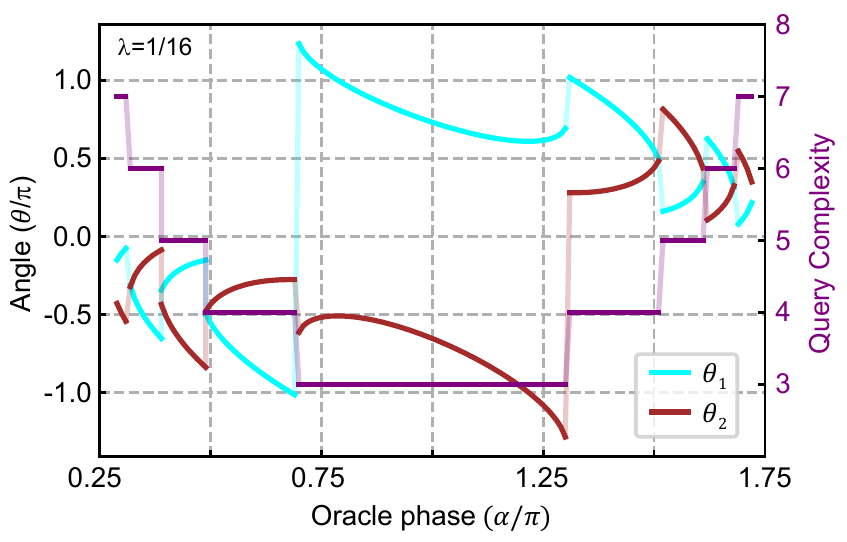}
	\caption{Numerically computed values of phase parameters $\{\theta_1, \theta_2\}$ (cyan and brown lines respectively) as a function of oracle phase $\alpha$ for $\lambda=2^{-4}$ to achieve zero failure rate. Note that any $\theta \mod 2\pi$ is a valid solution. The $\alpha$ values at which query complexity (purple line) jumps is dependent on the value of $\lambda$.}
	\label{fig:alpha}
\end{figure}
The D2p protocol can be applied to the cases when the oracle phase $\alpha\neq\pi$. While $k_{\rm opt}$ steps are sufficient to achieve zero failure rate for a range of $\alpha$ around $\pi$, increasingly more number of iterations are required with larger deviations. This property is demonstrated in Fig.~\ref{fig:alpha} for $\lambda=1/16$, chosen as an example. Exploring the dependence of optimal number of iterations and corresponding phase parameters as function of $\alpha$ and $\lambda$ is a subject of future research.

\bibliography{DGrover.bib}

\end{document}